\documentclass[english]{extarticle}
\usepackage[LGR,T1]{fontenc}
\usepackage[latin9]{inputenc}
\usepackage{geometry}
\usepackage{array}
\usepackage{verbatim}
\usepackage{units}
\usepackage{amstext}
\usepackage{graphicx}
\usepackage[numbers]{natbib}

\makeatletter

\DeclareRobustCommand{\greektext}{%
  \fontencoding{LGR}\selectfont\def\encodingdefault{LGR}}
\DeclareRobustCommand{\textgreek}[1]{\leavevmode{\greektext #1}}
\ProvideTextCommand{\~}{LGR}[1]{\char126#1}

\providecommand{\tabularnewline}{\\}

\@ifundefined{date}{}{\date{}}

\def\Institution#1{\begin{center} {\normalsize {\it #1} } \end{center}}

\makeatother

\usepackage{babel}
\begin{document}
\title{The REDTOP experiment}
\author{C. Gatto \\
On Behalf of REDTOP Collaboration}

\maketitle
\Institution{NFN Sezione di Napoli, via Cinthia, 80126 Napoli, Italy\\ also, Northern Illinois University, De Kalb (IL)} 
\begin{abstract}
The $\eta$ and $\eta'$ mesons are almost unique in the particle
universe since they are Goldstone and the dynamics of their decays
are strongly constrained. Because the $\eta$ has no charge, decays
that violate conservation laws can occur without interfering with
a corresponding current. The integrated $\eta$-meson samples collected
in earlier experiments amount to $\sim10^{9}$ events, dominated by
the WASA-at-Cosy experiment. A new experiment, REDTOP (Rare Eta Decays
with a TPC for Optical Photons), is being proposed, either at the
proton Delivery Ring of Fermilab or at the Proton Synchrotron of CERN,
with the intent of collecting in excess of 10$^{12}$ $\eta$/yr (10$^{10}$
$\eta'$/year) for studying rare decays. Such statistics are sufficient
for investigating several symmetry violations, and for searching particles
beyond the Standard Model. Recent studies have indicated that REDTOP
has very good sensitivity to processes that couple the Standard Model
to new physics through three of the so-called \emph{portals}: the
vector, the scalar and the axion portal. An upgraded version of the
detector will be proposed, at a later stage, for a run at the PIP-II
accelerator complex, where the $\eta$ mesons will be produced in
tagged mode. The physics program, the accelerator systems and the
detector for REDTOP are discussed.
\end{abstract}
\begin{quotation}
\emph{Talk presented at the 2019 Meeting of the Division of Particles
and Fields of the American Physical Society (DPF2019), July 29--August
2, 2019, Northeastern University, Boston, C1907293}.

\medskip{}
\end{quotation}

\section{Introduction}

\label{intro} In the last few years, the evidence that the Standard
Model of particle physics is incomplete has becoming more stringent
and the experimental community is pressed to engaging in new experiments
that would help to uncover the New Physics. The LHC complex was designed
to search for physics Beyond the Standard Model (BSM) at energies
of the order of few TeV. The failure, at least up to this writing,
in discovering any hint of it is an indication of at least two facts:
a) the New Physics most immediately accessible to the experimenter
probably lies at low energies, rather than at high energies, and b)
such New Physics is elusive, and it the couple to the Standard Model
physics too faintly to be detected by experiments at colliders. While
the latter, in fact, have several nice features from the experimental
point of view, the cannot compete in terms of luminosity with fixed
target experiments.

The latest theoretical models prefer gauge bosons in the MeV-GeV mass
range\citep{PhysRevD.80.095024,Reece_2009,Bjorken_2009}. In such
mass regime, strong astrophysical and cosmological constraints are
weakened or eliminated, while constraints from high energy colliders
are, in most cases, inapplicable. On the other side, in order to successfully
explore those models, beam luminosities considerably higher than those
available at colliders are required , as the former usually, do not
exceed the 10$^{34}$-10$^{35}$ cm$^{-2}$sec$^{-1}$ range. Fixed
target experiments in most cases exceed that range, as they can be
carried at high intensity facilities and under reasonably controlled
conditions (as compared, for example, to beam dump experiments).

\medskip{}

The design of the REDTOP experiment is based on the above observations.
It is a high intensity $\eta$/$\eta$'-factory in a fixed target
configuration with beam luminosity of at least 10$^{34}$ cm$^{-2}$sec$^{-1}$.
The mass range for potential discoveries is roughly {[}15 MeV-950
MeV{]}, limited on the lower side by the resolution of the detector
and on the upper side by the mass of the decaying mesons.

\section{Rationale for an $\eta/\eta'$-factory}

The $\eta$ and $\eta$ ' meson have been widely studied in the past,
as their special nature has attracted the curiosity of the scientists\citep{Nefkens_1996}.
The $\eta$ is a Goldstone boson, therefore its QCD dynamics is strongly
constrained by that property. In nature there are only few Goldstone
bosons. Furthermore, the $\eta$ is, at the same time, an eigenstate
of the \textbf{\textit{C}}\textit{, }\textbf{\textit{P}}\textit{,
}\textbf{\textit{CP, I}}\textit{ }and \textbf{\textit{G}} operators
with all zero eigenvalues (namely: $I^{G}J^{PC}=0^{+}0^{-+}$) which
makes it identical (except for parity) to the vacuum or the Higgs
boson. In that respect, it is a very pure state, carrying no Standard
Model charges and its decays do not involve charge-changing currents.
Any coupling to BSM states, therefore, does not interfere with Standard
Model charge changing operators. Another consequence of the above
properties is that the $\eta$ and $\eta$ ' have an unusually small
decay width, electromagnetic and strong decays are suppressed up to
the order O(10$^{-6}$ ) favoring the study of more rare decays, especially
those related to BSM particles and to violation of discrete symmetries.

\begin{figure}[!]
\begin{centering}
\includegraphics[scale=0.66]{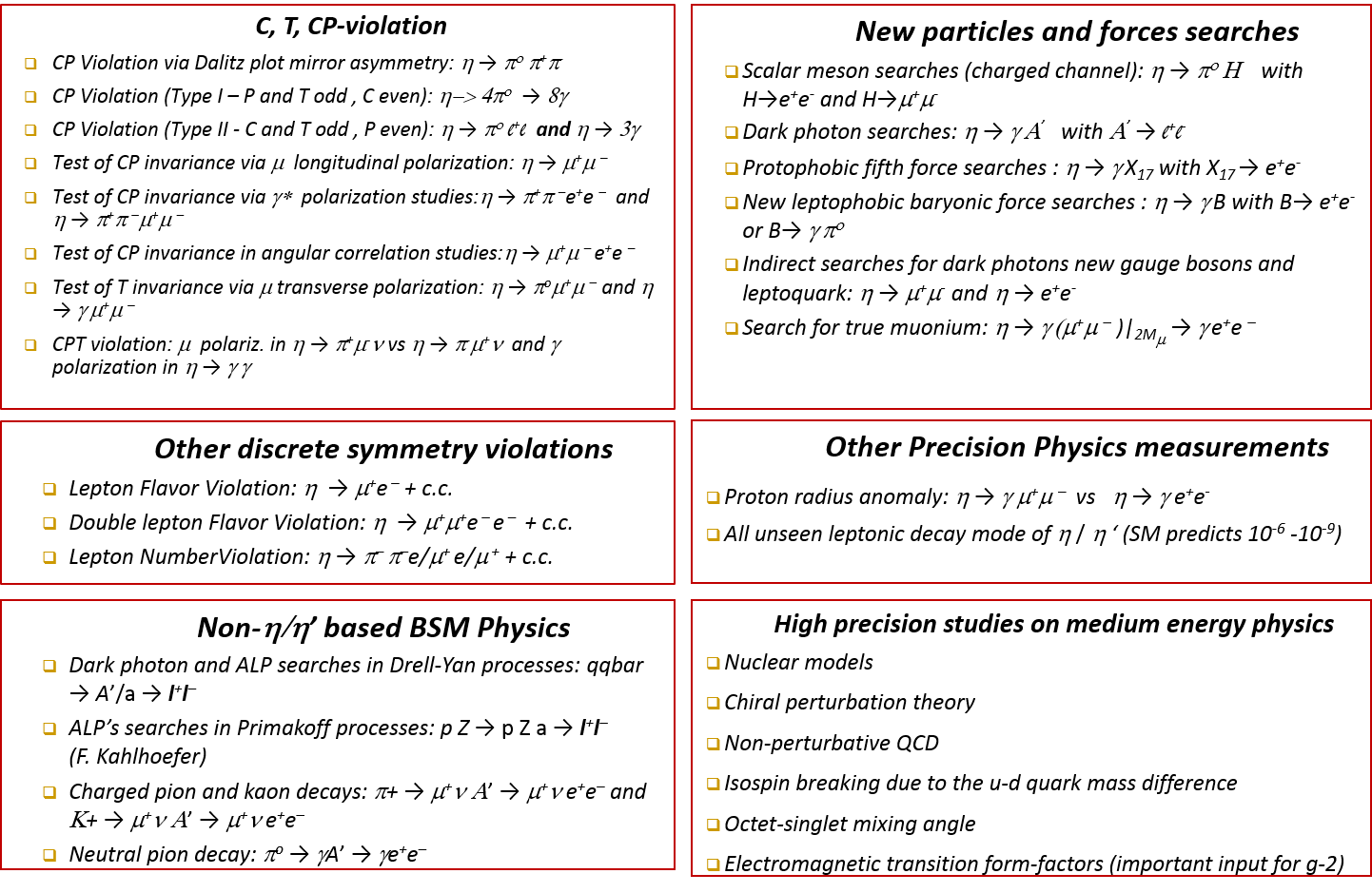} \caption{Physics landscape for an $\eta/\eta'$ factory.}
\par\end{centering}
\label{fig:physics_landscape}
\end{figure}

Considering the present limits on the parameters associated to BSM
physics and the practical limitations of current detector technologies,
the next generation of experiments should be designed with the goal
of producing no less than$10^{11}-10^{12}$ $\eta$ mesons. The physics
reach of an experiment with such statistics is quite broad, spanning
several aspects of physics BSM. A summary of the physics landscape
spanned by REDTOP is presented in Fig. \ref{fig:physics_landscape}
Along with BSM Physics, the availability of such a large sample of
flavor-conserving mesons will also allow to probe the isospin violating
sector of low energy QCD to an unprecedented degree of precision.
The large number of processes that could be studied at the proposed
$\eta/\eta'$-factory will provide not only a nice scientific laboratory
but also the source of many topics for Ph.D. thesis.

\medskip{}

The REDTOP experiment is currently being proposed for running either
at CERN or Fermilab. BNL is also a suitable laboratory. The design
luminosity in the former case, corresponds to the production of $\sim10^{12}$
$\eta$ mesons/year, being limited by the availability of an adequately
intense beam. The design luminosity if running at Fermilab, corresponds
to the production of $\sim10^{13}$ $\eta$ mesons/year, in this case
the limitation deriving from the detector and DAQ technologies available
in the foreseeable future. The production cross section for the $\eta'$
meson, in a similar kinematics regime, is about two order of magnitude
smaller, pointing to an expected meson yield of $10^{10}-10^{11}$
$\eta$' mesons/year.

\section{Golden channels}

Among the physics processes listed in Fig. \ref{fig:physics_landscape},
some have a larger signal/background ratio and have more potential
for a discovery or for a direct measurement. Some of these g\emph{olden
channels} are briefly discussed below.

\subsection{CP Violation from Dalitz plot mirror asymmetry in $\eta/\eta'$$\rightarrow\pi^{+}\pi^{-}\pi^{o}$}

Considering the $\eta$ meson decay:

\begin{equation}
\eta\rightarrow\pi^{+}\pi^{-}\pi^{o}\label{eta-tre-pi}
\end{equation}

if \textbf{CP} invariance is satisfied, the dynamics of the charged
pions should be totally symmetric. This would imply that, the Dalitz
plot of decay (\ref{eta-tre-pi}) should show no signs of asymmetries.
If on the other hand the Dalitz plot shows any mirror-asymmetry, this
would be an indication of \textbf{CP} and\textbf{ C} violation. This
asymmetry cannot be generated by SM operators at tree level, nor by
higher level operators that generate also EDMs (at tree level). Therefore,
CP-violation from this process  is not bounded by EDM as is the case,
for example, for the process: $\eta\rightarrow4\pi$. Consequently,
this channel is complementary to EDM searches even in the case of
T and P odd observables, since the flavor structure of the $\eta/\eta'$
is different from the nucleus. In this situation, CP violation could
arise from the interference of the CP-conserving weak interaction
with a new interaction that breaks C and CP generating a charge asymmetry
in the momentum distribution of the$\pi^{+}\pi^{-}$ systems. Recent
studies\citep{gardner2019patterns} indicate that the C- and CP-violating
amplitude with total isospin I=2 is much more severely suppressed
than that with total isospin I=0. Consequently, the I=0 $\eta/\eta'$
systems is privileged, compared to any other mesonic systems, for
CP-violation searches.

\begin{figure}[!]
\begin{centering}
\includegraphics{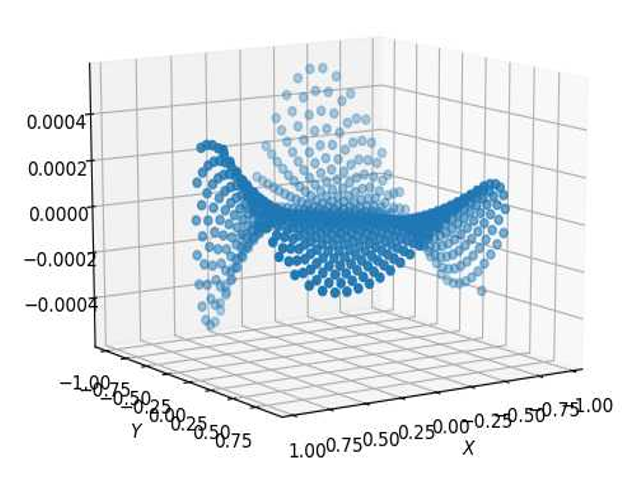} \caption{3-D representation of the Dalitz plot for the $\eta\rightarrow\pi^{+}\pi^{-}\pi^{o}$
process (courtesy of S. Gardner and J. Shi).}
\par\end{centering}
\label{fig:dalitz_plot}
\end{figure}

A 3-dimensional representation of the X and Y variables for the process
(\ref{eta-tre-pi}) is shown in Fig. \ref{fig:dalitz_plot}. The appearance
of terms that are odd in X would indicate both C and CP violation.
The detection of charged pions in REDTOP is based on the measurement
of the Cerenkov angle of the photons radiated in the aerogel (cfr.
Sec. \ref{subsec:The-Optical-TPC}). Therefore, the non-uniformity
of the magnetic field, which in general corresponds the largest contribution
to the systematic error in the asymmetry of kinematics variables of
positive and negative charged particles, plays no role at REDTOP.
The expected sensitivity for this process is currently under study
with the REDTOP detector, although it is expected to be higher than
with more traditional, magnetic spectrometers.

\subsection{Dark photon searches}

A search for a dark photon can be carried out at an $\eta/\eta'$-factory
by looking for final states with a photon and two leptons. Considering
the process:

\begin{equation}
\eta\rightarrow\gamma A'\rightarrow\gamma+l^{+}l^{-}\label{eq:dark_photon}
\end{equation}

in the hypothesis that the mass of this dark photon is smaller than
the mass of the $\eta/\eta'$ meson, it will be relatively straightforward
to observe it. A detailed simulation of the process and the expected
background, including many instrumental effects, has been performed
by the Collaboration within the ``Physics Beyond Collider'' program\citep{Alemany:2019vsk}.

\begin{figure}[!]

\subsection{\protect\includegraphics[scale=0.6]{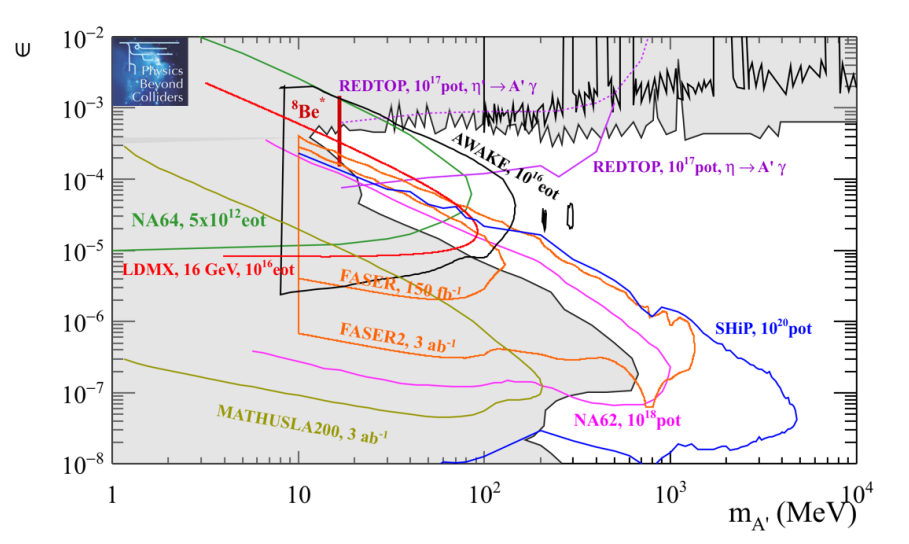} \protect\caption{REDTOP sensitivity to the $\epsilon$ parameter for 10$^{12}$ POT.}
}

\label{fig:dark_photon}
\end{figure}

An integrated beam flux of $10^{12}$ POT (as available at CERN, see
Sec. \ref{sec:The-Acceleration-Scheme} below) has been assumed. A
simple ``\emph{bump-hunt}'' analysis was performed, by looking at
the invariant mass of di-leptons associated to a prompt photon. The
sensitivity to the $\epsilon$ coupling constant for $\eta/\eta'$
combined runs is shown in Fig. \ref{eq:dark_photon}. The largest
contributing background was found to be from mis-measured leptons
from the 3-body decay $\eta\rightarrow\gamma+l^{+}l^{-}$. The sensitivity
to this channel would be greatly enhanced by including in the analysis
the reconstruction of secondary vertexes (associated to long decaying
dark photons) and with a larger integrated beam flux (available, for
example,at Fermilab or BNL). These studies are currently in progress.

\subsection{Searches for Axion Like Particles (ALP)}

This study has been performed as part of the ``Physics Beyond Collider''
program\citep{Alemany:2019vsk}, where specific assumptions were done
regarding the coupling of an ALP to the Standard Model. Among the
three benchmark processes designated by the PBC for this portal, two
of them have been exploited at REDTOP: a search for ALP with fermion
dominant coupling and with gluon dominant coupling.

Assuming a single ALP state which predominant couples to fermions,
all phenomenology (production and decay) can be determined as a function
of \{m$_{ALP}$, f$_{l}^{-1}$, f$_{q}^{-1}$\}. For this study, a
further assumption: f$_{l}^{-1}$= f$_{q}^{-1}$was made. The processes
considered are:
\begin{equation}
\eta\rightarrow\pi^{+}\pi^{-}ALP\rightarrow\pi^{+}\pi^{-}+l^{+}l^{-}\label{eq:ALPS_fermion}
\end{equation}
\begin{equation}
\eta\rightarrow\pi^{+}\pi^{-}ALP\rightarrow\pi^{o}\pi^{o}+l^{+}l^{-}\label{eq:ALPS_fermion-1}
\end{equation}

An integrated beam flux of $10^{12}$ POT (as available at CERN, see
Sec. \ref{sec:The-Acceleration-Scheme} below) has been assumed. A
simple ``\emph{bump-hunt}'' analysis was performed, looking at the
invariant mass of di-leptons associated to two pions. The sensitivity
to the $g_{Y}=\nicefrac{2v}{f_{l}}$ coupling constant for $\eta+\eta'$
combined runs is shown in Fig. \ref{eq:ALPS_fermion} (left). The
largest contributing background was found to be from mis-measured
particles from the 4-body decay $\eta\rightarrow\pi\pi l^{+}l^{-}$.

\begin{figure}[!]
\begin{centering}
\includegraphics[width=10cm]{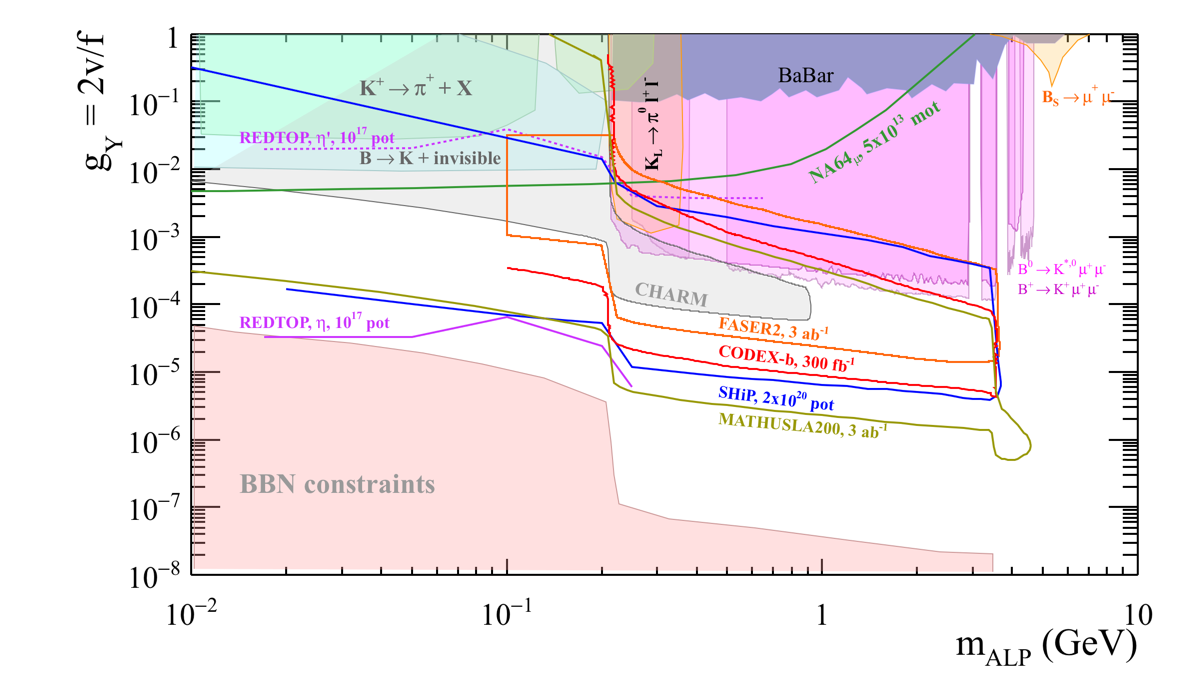} \includegraphics[width=10cm]{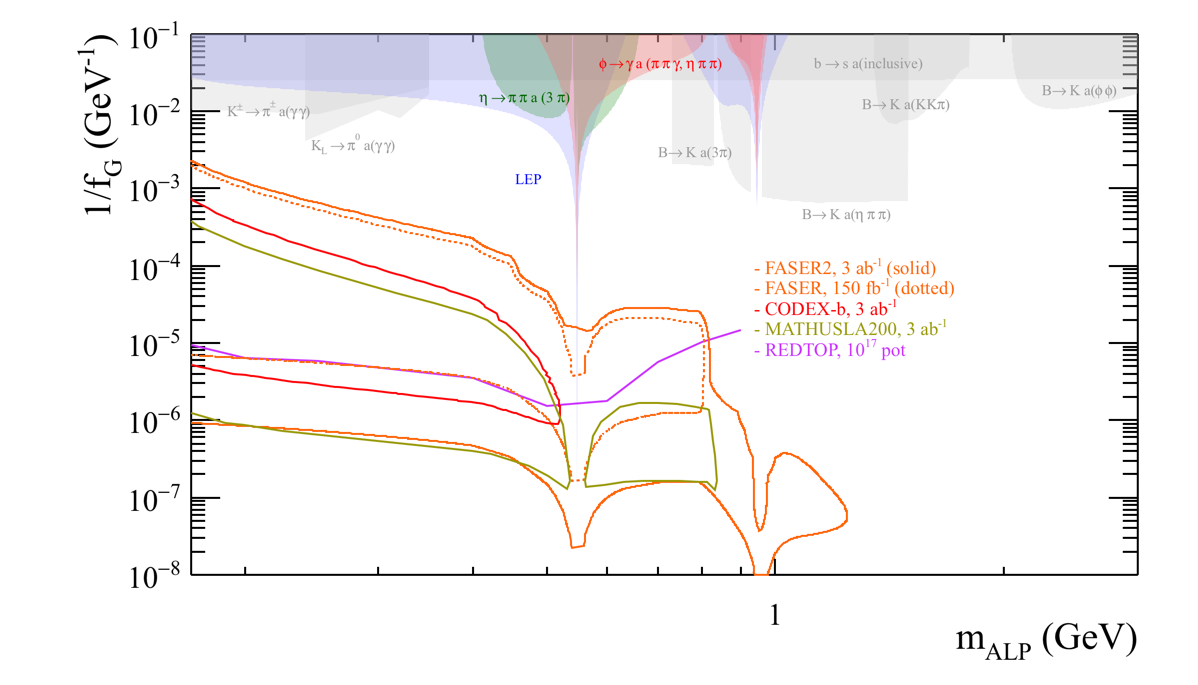}
\caption{REDTOP sensitivity for ALP's with dominant lepton (top) and gluon
(bottom) coupling for$10^{17}$ POT with $E_{kin}$=1.8 GeV on Be
targets.}
\par\end{centering}
\label{fig:pi_p_multiplicity-1}
\end{figure}

If the ALP has a dominant coupling to gluons, instead, it would mostly
be produced from beamsstrahlung processes associated to the proton
beam interacting with the target matter. Several processes can be
responsible for the emission of an ALP: a) Drell-Yan processes: $q\bar{q}\rightarrow ALP\rightarrow l^{+}l^{-}$,
b) proton beamsstrahlung processes\citep{BLUMLEIN2014320}: $p\,N\rightarrow p\,N\,ALP$
with $ALP\rightarrow l^{+}l^{-}$, c) Primakoff processes\citep{Dobrich2016}:
$p\,Z\rightarrow p\,Z\,ALP$. The sensitivity to the f$_{g}^{-1}$
parameter is shown in Fig. \ref{eq:ALPS_fermion} (right).

The sensitivity to a ALP decaying into leptons would be greatly enhanced
by including in the analysis the reconstruction of secondary vertexes
(observed in case of long decaying ALPs) and with a larger integrated
beam flux (available, for example,at Fermilab or BNL). These studies
are currently in progress.

\subsection{New scalar particles}

A scalar H could be observed in a $\eta/\eta'$ final state in association
with a $\pi^{0}$:

\begin{equation}
\eta\rightarrow\pi^{0}H'\rightarrow\pi^{0}+l^{+}l^{-}.\label{eq:scalar}
\end{equation}

Within the Standard Model, this process can only occur via a two-photon
exchange diagram with a branching ratio of the order of $10^{-9}$.
If such a light particle exists, even with a mass larger than the
$\eta/\eta'$ meson, which couples the leptons to the quarks, the
probability for this process could be increased by several orders
of magnitude, changing dramatically the dynamics of the process. Two
groups of theoretical models postulating a BSM light scalar are receiving
great attention lately: the Minimal Extension of the Standard Model
Scalar Sector\citep{PhysRevD.75.037701,PhysRevD.94.073009} and the
models containing Higgs bosons with large couplings to light quarks\citep{batell2018probing,Egana-Ugrinovic:2018znw,Egana-Ugrinovic:2019dqu}.
From the experimental point of view, these models are complementary:
the former predicting large coupling to the b-quark and to gluons
but a small one to the light quarks, while the latter predicts a large
coupling to light quarks. An observation at a $\eta/\eta'$ factory
of the process (\ref{eq:scalar}) would be an indication that the
second set of models would be the most likely extension to the SM.
Vice-versa, an observation of a scalar at a B-factory but not at REDTOP
would favor the first groupp of models.

A detailed simulation of the process (\ref{eq:scalar}) and of the
foreseen background, including many instrumental effects, has been
performed by the Collaboration within the ``Physics Beyond Collider''
program\citep{Alemany:2019vsk} assuming the Minimal Extension of
the Standard Model Scalar model.

\begin{figure}[!]

\subsection{\protect\includegraphics[scale=0.6]{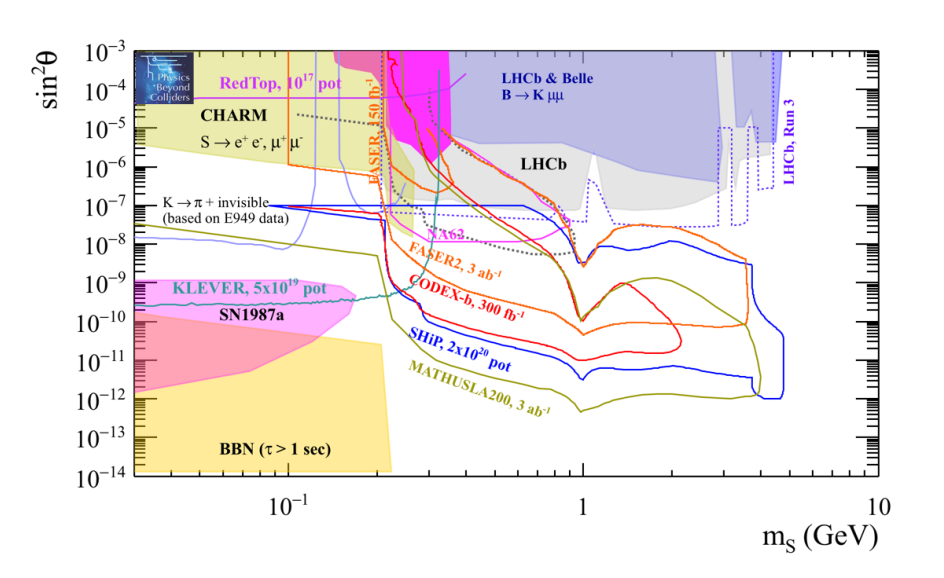} \protect\caption{REDTOP sensitivity to the $\sin^{2}\theta$ parameter\citep{PhysRevD.75.037701,PhysRevD.94.073009}
for 10$^{12}$ POT.}
}

\label{fig:dark_scalar}
\end{figure}

An integrated beam flux of $10^{12}$ POT (as available at CERN, see
Sec. \ref{sec:The-Acceleration-Scheme} below) has been assumed. A
simple ``\emph{bump-hunt}'' analysis was performed, looking at the
invariant mass of di-leptons associated to a prompt photon. The sensitivity
to the $\sin^{2}\theta$ parameter \citep{PhysRevD.75.037701,PhysRevD.94.073009}
is shown in Fig. \ref{fig:dark_scalar}. The largest contributing
background was found to be from the 3-body decay $\eta\rightarrow\gamma+l^{+}l^{-}$
where an extra $\gamma$ fakes a $\pi^{0}$ in the final state.

A preliminary sensitivity analysis for several experiments (including
REDTOP) based on the second set of models can be found in Ref. \citep{batell2018probing}

\subsection{Other Golden channels}

Besides the rare processes discussed above, several other processes
have a good potential for discovery. They are briefly listed below.

\subsubsection{Test of CP invariance via $\gamma${*} polarization studies:}

CP-violation can also be investigated with a virtual photon decaying
into a lepton-antilepton pair, as in:

\begin{equation}
\eta\rightarrow\pi^{+}\pi^{-}\gamma^{*}\text{ with }\gamma^{*}\rightarrow l^{+}+l^{-}.
\end{equation}

by considering the asymmetry

\begin{equation}
A_{\Phi}=\frac{N(\sin\phi\cos\phi>0)-N(\sin\phi\cos\phi<0)}{N(\sin\phi\cos\phi>0)+N(\sin\phi\cos\phi<0)}
\end{equation}

where $\phi$ is the angle between the decay planes of the lepton-antilpton
pair and the two charged pions. CP invariance requires $A_{\Phi}$
to vanish. At the present, the measurement of such asymmetry performed
by the WASA collaboration \citet{Adlarson-15} is the best available,
and it is consistent with zero within the measurement errors. Unfortunately,
that measurement is largely dominated by the statistical error, from
the production of only $\sim10^{9}\:\eta$-mesons. The larger statistics
of REDTOP will improve the systematic error by almost two orders of
magnitude.

\subsubsection{The proton radius anomaly}

One of the existing and still unexplained anomalies present in the
Standard Model is related to the measurements of the proton radius
$R_{p}$ with electron and muon probes\footnote{Although a very recent measurement has reconciled the anomaly, the
difference between the averages is still an open question. REDTOP
will contribute independently to clarify the situation.}

The processes involved are:

\begin{equation}
\eta\rightarrow\gamma e^{+}e^{-}\text{ and }\eta\rightarrow\gamma\mu^{+}\mu^{-}\label{eq:proton_radius}
\end{equation}

\begin{figure}[!htbp]
\centering{}\includegraphics[width=0.65\columnwidth]{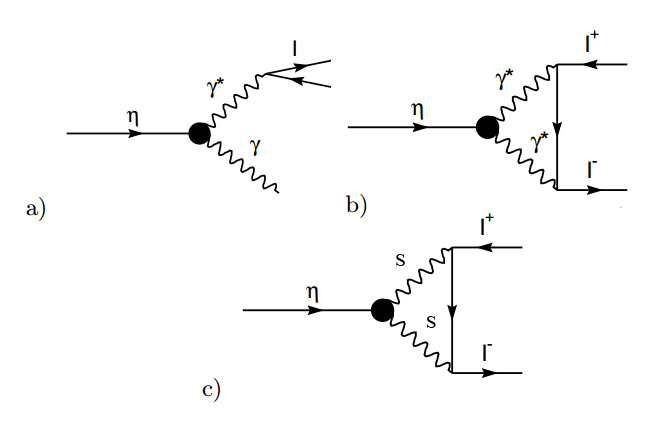} \caption{\label{fig:rp} Diagrams contributing to the proton radius}
\end{figure}

Other types of $R_{p}$ measurements, using muonic atoms (see, for
example, CODATA-2012), or using elastic scattering of electrons and
muons on hydrogen atoms, have found a discrepancy corresponding to
several sigma in the electron vs muon case. It is worth noting that
such processes occur mainly trough the exchange of one virtual photon,
while processes of the type $\eta\rightarrow\gamma l^{+}l^{-}$ can
occur either via one photon, as shown in figure \ref{fig:rp} diagram
a), or two photons, as shown in figure \ref{fig:rp} diagram b).

A light scalar particle S with a different coupling with electrons
and muons would mediate this process, as shown in diagram c) of figure
\ref{fig:rp}, would explain this anomaly of the Standard Model. Therefore,
an experiment able to precisely measure the branching ratios of this
particle might help in explaining the $R_{p}$ anomaly.

In REDTOP, the processes (\ref{eq:proton_radius}) are detected simultaneously
and within the same experimental apparatus. Consequently, most of
the systematic errors are common to the two processes and they factor
out in the ratio of the corresponding branching ratios, enhancing
the precision of the overall measurement.

\section{\label{sec:The-Experimental-Technique}The Experimental Technique
and Requirements}

The most efficient way to produce $\eta/\eta'$ mesons is by hadro-production
from nuclear scattering of a proton beam onto a nuclear target. Above
E$_{kin}$$\approx1.4$ GeV several intra-nuclear baryonic resonances
are created, whose decay produce an $\eta$-meson. The $\eta$-meson
production cross section increases almost linearly above such threshold,
while the QCD background remains approximately constant. Although
an increase of the beam energy would impact the $\eta$ yield positively,
the larger multiplicity of the background and the momentum of the
final state particles (mostly, baryons, slow pions and nuclear remnants)
would complicate the event selection and affect adversely the\emph{
signal/noise} ratio.

\subsubsection*{Beam and target requirements}

Extensive studies have been conducted with the \emph{GenieHad}\citep{GenieHad_2012}
event-generator framework using multiple beam and target parameters.
Such studies indicate that a proton beam with E$_{kin}$$\approx1.8-1.9$
GeV impinging on a thin, low-Z, high-$\nicefrac{A}{Z}$ material like
a Beryllium or Lithium target has been found as an optimal configuration.
The low-Z requirement helps in minimizing the multiplicity of the
fragmentation products, QCD-generated. For example, the multiplicity
of primary neutrons produced by a Nb target increases more than 20\%
wrt a Be target, while the increase of primary protons is almost 90\%
(from \emph{GenieHad} simulations). The high-$\nicefrac{A}{Z}$ requirement
reflects the factor $\sim$7x between the $\eta$ production cross
section for neutrons vs protons. Finally, the thin targets requirement
constrains, on one side, the z-coordinate of the $\eta$ production
vertex, and, on the other side, it minimizes the multiple scattering
affecting the $\eta$ decay products when they escape from the target.
By using 10 foils, spaced 10 \emph{cm} apart, one would provide the
necessary total material budget for the required luminosity, while
the pile-up due to multiple beam interactions within the same trigger
could be easily reduced with vertexing techniques. A similar study
for the $\eta'$ case yields an optimal range for E$_{kin}$ of about
3.5-4 GeV.

The expected inelastic interaction rate for the combination of energy
range and target systems indicated above as well as the $\eta/\eta'$
production has been calculated with several nuclear scattering models
available in \emph{\citep{GenieHad_2012}} (more specifically, \emph{Urqmd,
Gibuu, PHSD }and\emph{ Jam}). They produced comparable results and
predict an inelastic interaction probability of the order of 1\% with
a probability for an event to contain a $\eta$ meson of the order
of 0.25\% - 0.5\%. The probability to generate an $\eta'$ meson is
two orders of magnitude smaller. A CW proton beam delivering $1\times10{}^{11}$POT/sec
would generate a rate of inelastic interaction of about 1 GHz and
a $\eta$-meson yield of $2.5-5\times10{}^{6}$$\eta/sec$, corresponding
to $2.5-5\times10{}^{13}$$\eta/yr$. The beam power corresponding
to the above parameters is approx. 30 W, of which 1\% (or 300 mW)
is absorbed in the target systems. Each of the 10 foils will have
to dissipate about 30 mW, which is easily achievable radiatively.
These number roughly double for the beam required at a $\eta'$-factory.

\subsubsection*{Detector requirements}

\begin{figure}[!]
\begin{centering}
\includegraphics[scale=0.45]{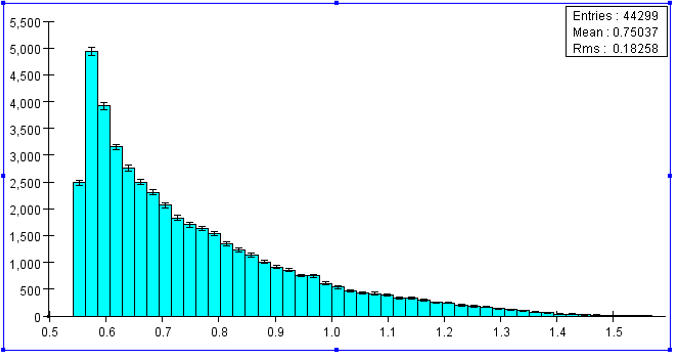} \caption{Total energy of $\eta$ mesons from a $E_{kin}$=1.8 GeV proton beam
scattered on REDTOP targets (\emph{GenieHad} simulation).}
\par\end{centering}
\label{fig:eta_energy}
\end{figure}

The $\eta$ mesons generated with the above beam parameters are almost
at rest in the lab frame, receiving only a small boost in the direction
of the incoming beam (see Fig. \ref{fig:eta_energy}). This is a consequence
of their production mechanism as they are the decay products of an
intranuclear resonance in the target material. Studies performed with
\emph{GenieHad} have indicated that the $\eta-$spectrum is almost
independent from the beam energies in the {[}1.46-2.1{]} GeV range.
Consequently, an hermetic detector covering the entire solid angle,
is one of the requirements for REDTOP.

A correct identification of the final states particles is, also, of
paramount importance for an experiment exploring rare processes. On
one side, the vast majority of the processes listed in Fig. \ref{fig:physics_landscape}
have two leptons at at least one photon in the final state. On the
other side, the QCD background, which has a production rate two order
of magnitudes larger, is populated almost invariably by protons, neutrons,
slow pions and nuclear remnants. These are relatively slow particles,
as shown in Fig. \ref{fig:eta_energy} for the case of protons an
pions. Furthermore, neutrons impinging onto the calorimeter would
easily be mistaken as photons if a proper particle identification
(PID) is not in place. Finally, the identification of$e^{+}e^{-}$
pairs from photosn converting in the material upstream the central
tracker will help to reject the large background originating from
QCD production of $\pi{}^{o}$ particles. The detector is inserted
in a solenoidal magnetic field which bends the charged particles for
proper $P_{t}$ and electric charge measurement.

\begin{figure}[!]
\begin{centering}
\includegraphics[scale=0.58]{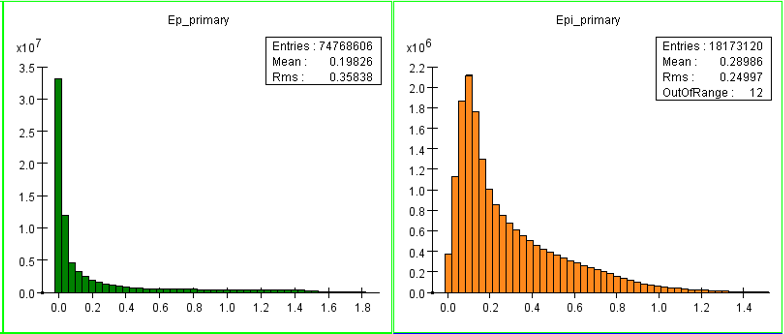} \caption{Momentum of protons (left) and charged pions(right) from inelastic
scattering of a $E_{kin}$=1.8 GeV proton on REDTOP targets (\emph{GenieHad}
simulation).}
\par\end{centering}
\label{fig:bkg_p_pi-1}
\end{figure}

\begin{figure}[!]
\begin{centering}
\includegraphics[scale=0.9]{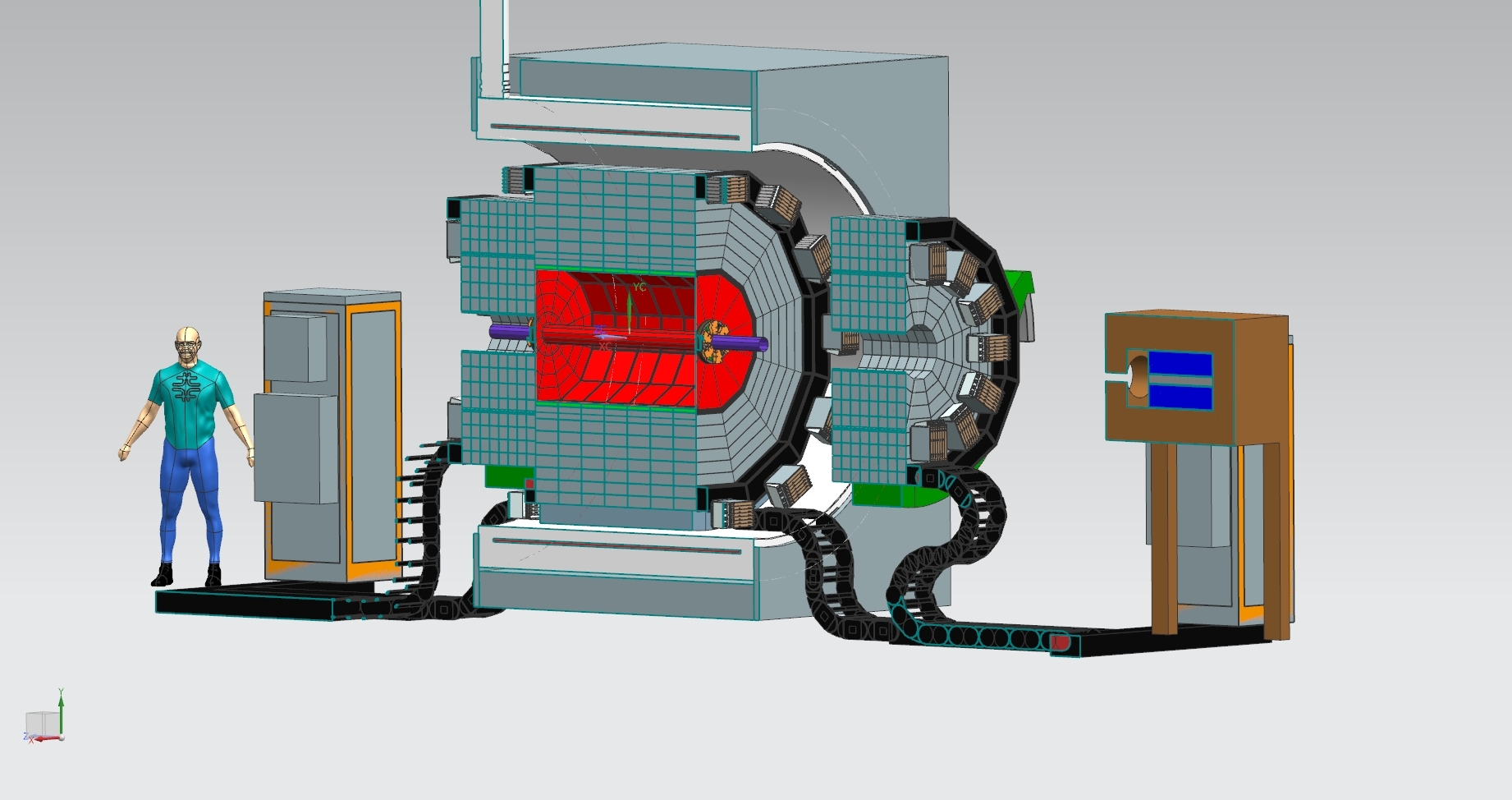} \caption{Cross section of REDTOP detector}
\par\end{centering}
\label{fig:detector-sketch}
\end{figure}

\subsubsection*{\label{subsec:The-Optical-TPC}The Optical TPC}

Protons and slow pions account for the largest charged background
at REDTOP. In average 0.6 charged pions and 1.8 protons are produced
every nsec in the target systems by the $E_{kin}$=1.8 GeV proton
beam.

\begin{figure}[!]
\includegraphics[width=8cm]{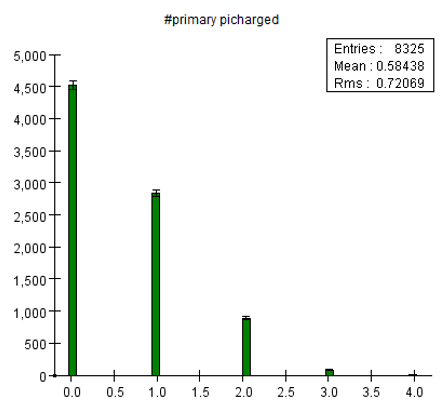} \includegraphics[width=8cm]{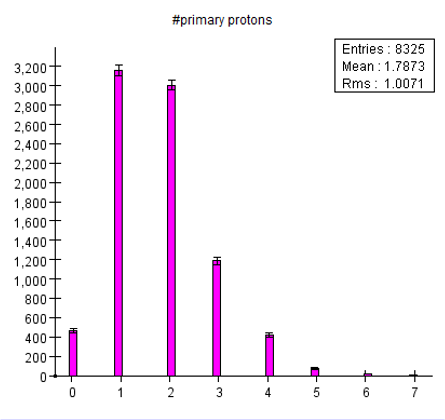}
\caption{Charged pions (left) and proton (right) multiplicity from inelastic
scattering of a $E_{kin}$=1.8 GeV proton on REDTOP targets (\emph{GenieHad}
simulation).}

\label{fig:pi_p_multiplicity}
\end{figure}
A charged track detector mostly blind to them is required in order
to identify more rare decays. These particles have a consistently
low value of $\beta.$ Consequently, the threshold characteristic
of the Cerenkov effect could be exploited to build a detector insensitive
to this hind of background.

The central tracker proposed for REDTOP (the red dodecagon in FIG.
\ref{fig:detector-sketch}) is an Optical TPC\citep{T1059,Oberta}
(OTPC). It measures the transverse momentum of electrons or the $\beta$
of heavier particles (muons and fast pions). It is a vessel filled
with gas ($CH_{4}$) with the inner wall covered by an aerogel radiator
with a thickness of 3 cm. The external walls of the chamber are instrumented
with optical photo-sensors (LAPPD or sipm's) with a pixel size of
few $mm^{2}$. When an electron or a positron with a $\beta$ above
the Cerenkov threshold of $CH_{4}$ traverses the gas, it bends in
the solenoidal magnetic field while also radiating photons. The sweep
of the latter, being collected by the photo-sensors at the external
walls of the device, has a specific pattern corresponding to the $P_{t}$
of the particle. Position and direction of the particle are obtained
by combining the pattern of the ring radiated when the aerogel is
crossed with the information of the impinging point on the photo-sensors.

Slower particles like muons and pions, typically with $\beta$ below
the Cerenkov threshold in the gas, produce a characteristic photon
ring when they cross the aerogel. This is detected from the OTPC photo-sensors
and reconstructed, providing a measurement of the kinematic parameters
of the particle. Protons and pions from QCD scattering are consistently
below detection sensitivity of the OTPC and do not constitute a sizable
background for the experiment. The magnetic field (0.6 T) also provides
the magnetic bottling preventing these very slow particles particles
from reaching the calorimeter with elevated rates.

\subsubsection*{ADRIANO2 calorimeter}

Dual-readout calorimetry\citep{DREAM_2000,LCWS2015} is a proposed
technique to achieve energy compensation in hadronic showers produced
in high energy experiments

. One particular implementation: ADRIANO, where both readout components
are integrally active, has been developed in recent years by T1015
Collaboration\citep{Gatto_2015_1,Gatto_2015_2}. One of the advantages
with ADRIANO is that, with a proper layout, the dual-readout method
can be used also for an electromagnetic calorimeter in a way that
the two read-out components would provides the PID of the impinging
particle. The application of dual-readout for PID purposes is illustrated
in Fig. \ref{fig:Plot-of-S} for particles of different species and
same energy (100 MeV). The separation between photons and neutrons,
is better than 4$\sigma$, and it exceeding the requirement of 99\%
established with software simulations.
\begin{center}
\begin{figure}
\centering{}\includegraphics[scale=0.55]{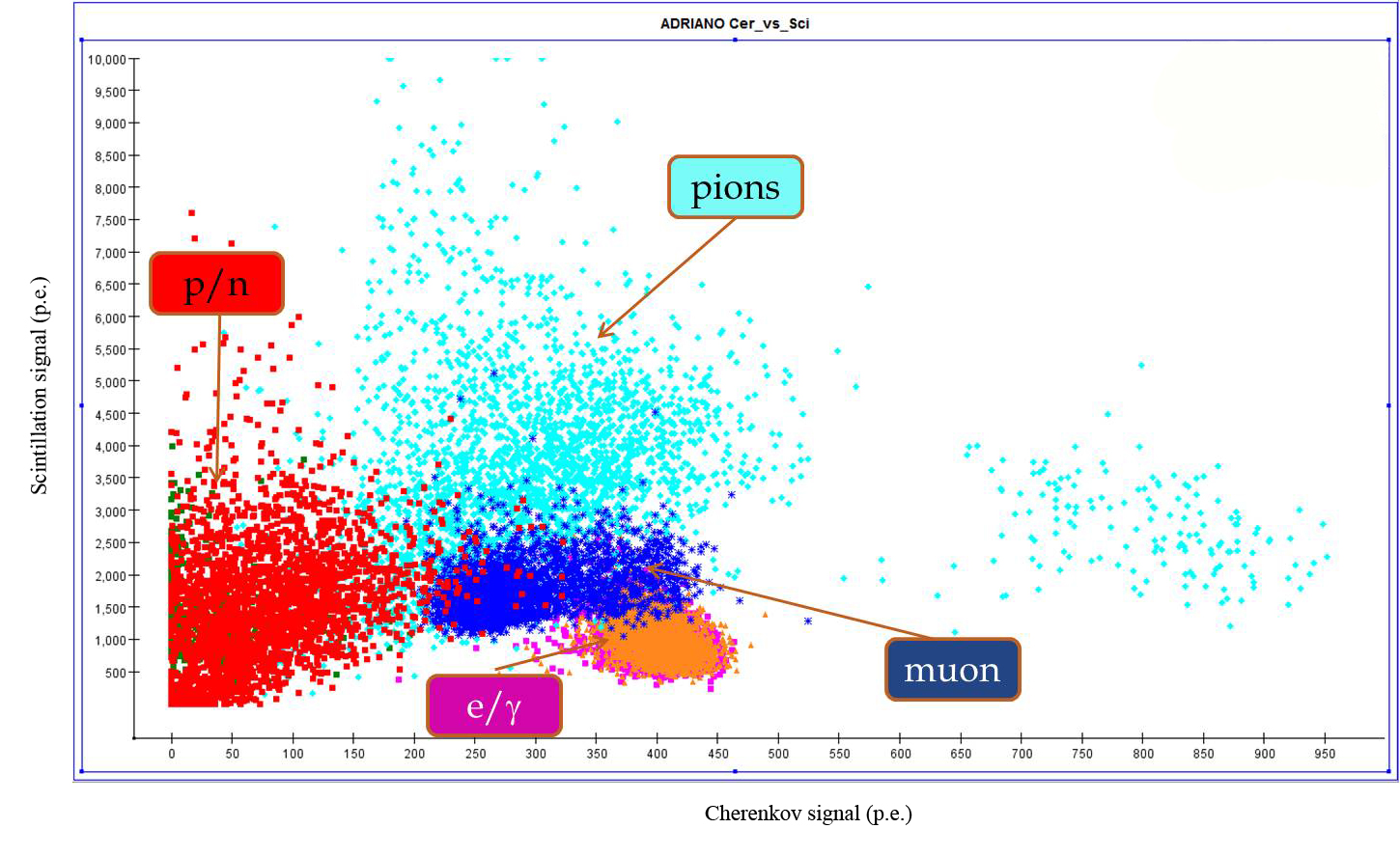}\caption{\label{fig:Plot-of-S}Plot of S vs \v{C} signals for several 100 MeV
particles in ADRIANO (ilcroot simulation)}
\end{figure}
\par\end{center}

REDTOP also requires a finer granularity than that achievable with
the, log based, ADRIANO technique. A newer implementation of ADRIANO
is being proposed: ADRIANO2, where the lead-glass and the scintillating
strips, read by WLS fibers, are replaced by tiles with direct sipm
readout\citep{TILES2009,TILES2011,MALLA_2019}.

\subsubsection*{Fiber tracker}

A fiber tracker, with an identical technology as implemented for the
LHCb upgrade\citep{Kirn_2017}, is located around the beam pipe, at
a radius r$\simeq$5 cm around the z-axis. Three superlayers are stacked
together, interspaced with structural foam of 1 cm providing mechanical
stability. Each superlayer is obtained from a mat of 5 layers of scintillating
fibers, with a diameter of 250 $\mu m$. The space resolution of each
superlayer is $\sim70\mu m$, as measured by LHCb at a test beam.
The fiber tracker serves two main purposes: a) it provides a vertex
measurement for the reconstructed event, and b) it identifies photons
converting into $e^{+}e^{-}$ pairs inside the OTPC (typically, in
the inner wall of the chamber or in the aerogel). The determination
of the primary vertex of the event identifies the foil where the inelastic
interaction occurred and it provides the position of the $\eta$-mesons.
If detached vertexes are found not belonging to the very few produced
$K_{short}$, that will be a clear indication that new physics is
being observed.

\subsubsection*{Superconducting solenoid}

A 0.6 T solenoidal magnetic field is required to measure the $P_{t}$
of the particles with $\beta\approx1$. The field will also magnetically
bottle the very low momentum particles, preventing them from reaching
the calorimeter. The solenoid built for the, now dismantled, Finuda
experiments\citep{Bert997} matches all operational and dimensional
parameter required for REDTOP. It fits nicely trough the hatch that
connects the AP50 hall (that would house REDTOP if running at Fermilab)
and it can be transported on rails to its final location. The Collaboration
has made a formal request to Istituto Nazionale di Fisica Nucleare
(owning the magnet) for a loan to the experiment.

\section{\label{sec:The-Acceleration-Scheme}The Acceleration Scheme}

Although the beam requirements for REDTOP are modest, none of the
existing HEP laboratories worldwide as a ready-to-go accelerator satisfying
all the conditions. The Collaboration has engaged in a broad exploration
and three possible hosting laboratories have been identified.

\paragraph*{Fermilab configuration.}

The low energy proton beam available at Fermilab has a pulsed structure
with 84 bunches and a fixed energy of 8 GeV. The buckets are distributed
to the accelerator complex and either accelerated or dumped in order
to create a secondary beam. On the other hand, REDTOP requires a Continuous
Wave (CW) proton beam with a user selectable energy, with a range
considerably lower than that available (cfr. Sec. \ref{sec:The-Experimental-Technique}).
The accelerator scheme\citep{Decel_2016} proposed for REDTOP foresee
the extraction of a single pulse from the booster (consisting of $\sim$4$\times$10$^{12}$
protons) which is subsequently injected in the Delivery Ring (former
debuncher in anti-proton production at Tevatron). The energy is, then
removed from the beam by operating the RF cavities in reverse, until
it reaches the required value. The time required for reaching 1.8
Gev is $\sim$5 seconds (about 3.3 for the 3.5 GeV case). Once the
desired energy is reached, the beam is kept circulating inside the
ring where the buckets relax adiabatically. Slow extraction to REDTOP
(located in the nearby AP50 hall) occur over $\sim$40 seconds. The
proposed accelerator scheme requires minimal changes to the existing
complex and involves no extra beam elements. In fact, the extra RF
cavity necessary to stabilize the energy removal process is already
available as a spare for the Mu2e experiment. Furthermore, the extraction
point to AP50 corresponds to a betatron phase advance of 270$^{o}$
compared to the location of the existing Mu2e electrostatic septum.
Consequently, the latter can be used also to shift the protons for
the final extraction via a Lambertson magnet. The total time to decelerate-debunch-extract
the beam is 51 sec, corresponding to a duty cycle $\sim$80\%.

With the above scheme, large beam losses in the Delivery Ring will
occur if beam is decelerated from injection at 8 GeV (corresponding
to $\gamma$=9.53) to <2 GeV ($\gamma$ = 3.13) through the natural
transition energy of the ring: $ $$\gamma{}_{t}$ = 7.64. Transition
can, however, be avoided by using select quadrupole triplets to boost
$\gamma{}_{t}$ above beam $\gamma$ by 0.5 units throughout deceleration
until $\gamma{}_{t}$ = 7.64 and beam $\gamma$ = 7.14 (corresponding
to E$_{kin}$=5.76 GeV ). Below 5.76 GeV the Delivery Ring lattice
is reverted to the nominal design configuration. Optical perturbations
are localized within each triplet while the straight sections are
unaffected thereby keeping the nominal M3 injection beamline tune
valid\citep{Decel_2016}.

\paragraph*{CERN configuration.}

A preliminary study has been performed at CERN to provide a beam with
REDTOP's requirements. Several possible schemes have been considered,
the most promising corresponding to the extraction of a 1.8 GeV beam
from the ProtoSynchrotron (PS) which could then be delivered to the
East Hall, where the experiment could be housed. For the moment a
24 GeV/c proton beam is routinely slow-extracted into the CHARM and
IRRAD facilities along the T8 beam line with a maximum intensity of
6.5$\times$10$^{11}$ protons per extraction over 0.4 seconds. REDTOP
would require a much longer flat top of e.g. 10 seconds at $\sim$1.8
GeV kinetic energy (as it can be delivered directly from the PS Booster)
in a cycle of 9 basic periods(10.8 s). No show-stoppers have been
identified up to date, although more studies are required. However,
the PS the duty cycle cannot be much higher than 50\%, which would
already have a significant impact on the existing CERN physics program.
Consequently, the expected luminosity available to REDTOP at CERN
is expected to be $\sim\nicefrac{1}{10}$ of that available at Fermilab.

\paragraph*{BNL configuration.}

AT BNL the REDTOP detector could sit comfortably at the end of the
existing C4 extraction line. Logistics would be optimal, since the
experimental hall is well accessible and serviced. The beam would
be slowly extracted from the AGS using well known techniques. The
instrumentation electronics of the extraction line is no longer functional
and it needs to be refurbished at a very modest cost ($\mathcal{O}$(100K\$)).
Altogether, BNL is an excellent candidate for hosting REDTOP.

\section{Tagged $\eta$-factory}

An upgraded version of the REDTOP, t-REDTOP, could be run in a later
stage of the experiment at the PIP-II facility, currently under construction
at Fermilab. At the 800 MeV, high intensity (100KW-1MW), CW proton
beam, the production mechanism of the $\eta$-meson is substantially
different. The nuclear process providing the $\eta$-meson would be:

\begin{equation}
p+De\rightarrow\eta+^{3}He^{+}\label{eq:tagged_eta}
\end{equation}

Therefore, t-REDTOP requires a gaseous Deuterium target and an extra
detector to tag the $^{3}He^{+}$ ion. The production cross section
for the process (\ref{eq:tagged_eta}) is approximately five orders
of magnitude smaller than at 1.8 GeV and the gaseous target reduces
further the luminosity of the beam. However, the large intensity of
PIP-II more than compensate for that and the number of $\eta-mesons$
produced at t-REDTOP is expected to be in the range {[}$10^{13}/year-10^{14}/year${]}.
The biggest advantages of a tagged $\eta$-factory are the following:
\begin{itemize}
\item By tagging the production of the $\eta$-meson via the detection of
the $^{3}He^{+}$ ion, the combinatorics background from $non-\eta$
events, is greatly diminished. Consequently, the sensitivity of the
experiment to New Physics is increased by a factor proportional to
the square root of that reduction;
\item By measuring the momentum of the $^{3}He^{+}$ ion in process (\ref{eq:tagged_eta}),
the kinematics of the reaction is fully closed. That portends to a
better measurements of the kinematics of the particles detected, since
a 4-C kinematic fit could be applied;
\item Since the kinematics is closed, any long lived, dark particle escaping
detection could be identified using the \emph{missing 4-momentum}
technique. The latter is considerably more powerful than the, 1-,
missing p$_{t}$ or missing energy proposed by some recent experiments
searching for dark matter.
\end{itemize}
The disadvantages of a tagged-$\eta$ experiment are due mostly to
the larger complexity of the experiment, requiring an extra detector
at very small angles and to the necessity to control the halo of the
beam at a much higher level. Also, the mass of new physics explored
will be lower, since no $\eta'$-meson could be produced at the PIP-II
design energy.

\section{Timeline}

The design of REDTOP experiment introduces several innovations from
the point of view of the accelerator as well as the detector. The
way the OTPC is operated at REDTOP, differs substantially from the
current applications\citep{T1059,Oberta}. Although, in terms of construction,
it is nothing more than a vessel filled with gas and surrounded by
photo-sensors, it still requires substantial R\&D before its design
is finalized.

A dual-readout has undergone extensive R\&D in the past 15 years both
in a sampling and integrally active configuration. The ADRIANO2 layout,
has been proposed recently and a preliminary R\&D has is currently
on-going within the Collaboration\citep{MALLA_2019}.

Finally, although no stopping issues have been identified for the
accelerator scheme at any of the hosting laboratories being considered,
further studies are necessary, which also include the full design
of the extraction line.

In summary, at least 2-3 years of intense R\&D are foreseen before
the final design of the experiment could obtained. Mechanical engineering
would require approximately 1-2 years, although it can occur concurrently
with the R\&D on the detector. Once the design is finalized, construction
and civil engineering would take between two and three years, although
the fabrication of the individual subdetectors could happen at different
times. The in-kind contribution of the existing Finuda solenoid would
help in reducing the construction schedules. Altogether, once approved,
the completion time of the experiment is of the order of 6-7 years.
After an engineering and a calibration run, the Collaboration intends
to take data for about one year at E=1.8 GeV and another year at E=3.5
GeV.

\section{Costing}

\subsection{Summary of costs}

A preliminary cost estimate\citep{ESPP_2018} for the construction
of the detector has been submitted to CERN in 2018 in partial fulfillment
of the requirements for participating to the European Strategy for
Particle Physics. Table \ref{tab:Preliminary-cost-estimate} summarizes
the costs estimate of the individual components. The total expected
cost, including a contingency factor of 50\%, is about \$ 51M. A cost
estimate, based on more stringent rules adopted in the US, is currently
under way.

\begin{table}
\centering{}%
\begin{tabular*}{12cm}{@{\extracolsep{\fill}}|>{\raggedright}p{8cm}|l|}
\hline 
\textbf{Solenoid} & \textbf{0.2}\tabularnewline
\hline 
\hline 
Refurbishing, shipping & 0.2\tabularnewline
\hline 
\noalign{\vskip0.2cm}
\end{tabular*}\medskip{}
\begin{tabular*}{12cm}{@{\extracolsep{\fill}}|>{\raggedright}p{8cm}|c|}
\hline 
\textbf{Supporting structure} & \textbf{1.0}\tabularnewline
\hline 
\noalign{\vskip0.2cm}
\end{tabular*}\medskip{}
\begin{tabular*}{12cm}{@{\extracolsep{\fill}}|>{\raggedright}p{8cm}|c|}
\hline 
\textbf{Target+beam pipe} & \textbf{0.5}\tabularnewline
\hline 
\noalign{\vskip0.2cm}
\end{tabular*}\medskip{}
\begin{tabular*}{12cm}{@{\extracolsep{\fill}}|>{\raggedright}p{8cm}|c|}
\hline 
\textbf{Fiber tracker} & \textbf{0.93}\tabularnewline
\hline 
\hline 
Fiber mats & 0.01\tabularnewline
\hline 
Tooling & 0.45\tabularnewline
\hline 
SiPM array & 0.1\tabularnewline
\hline 
Front-end electronics & 0.12\tabularnewline
\hline 
Back-end electronics & 0.05\tabularnewline
\hline 
Mechanics and cooling & 0.2\tabularnewline
\hline 
\noalign{\vskip0.2cm}
\end{tabular*}\medskip{}
\begin{tabular*}{12cm}{@{\extracolsep{\fill}}|>{\raggedright}p{8cm}|c|}
\hline 
\textbf{Optical-TPC} & \textbf{10.0}\tabularnewline
\hline 
\hline 
Vessel & 0.5\tabularnewline
\hline 
Aerogel & 1.0\tabularnewline
\hline 
Photo-sensors (LAPPD option) & 6.0\tabularnewline
\hline 
Front-end electronics & 1.8\tabularnewline
\hline 
Back-end electronics & 0.7\tabularnewline
\hline 
\noalign{\vskip0.2cm}
\end{tabular*}\medskip{}
\begin{tabular*}{12cm}{@{\extracolsep{\fill}}|>{\raggedright}p{8cm}|c|}
\hline 
\textbf{ADRIANO2} & \textbf{16.0}\tabularnewline
\hline 
\hline 
Pb-glass & 2.7\tabularnewline
\hline 
Cast scintillator & 0.75\tabularnewline
\hline 
Tile fabrication & 0.6\tabularnewline
\hline 
SiPM & 6.0\tabularnewline
\hline 
Front-end electronics & 4.0\tabularnewline
\hline 
Back-end electronics & 1.5\tabularnewline
\hline 
Mechanics and cooling & 0.5\tabularnewline
\hline 
\noalign{\vskip0.2cm}
\end{tabular*}\medskip{}
\begin{tabular*}{12cm}{@{\extracolsep{\fill}}|>{\raggedright}p{8cm}|c|}
\hline 
\textbf{Trigger} & \textbf{1.2}\tabularnewline
\hline 
\hline 
L0 + L1 & 1.0\tabularnewline
\hline 
L2 farm + networking & 0.2\tabularnewline
\hline 
\noalign{\vskip0.2cm}
\end{tabular*}\medskip{}
\begin{tabular*}{12cm}{@{\extracolsep{\fill}}|>{\raggedright}p{8cm}|c|}
\hline 
\textbf{DAQ} & \textbf{5.0}\tabularnewline
\hline 
\hline 
Digitizer & \tabularnewline
\hline 
Networking & \tabularnewline
\hline 
\noalign{\vskip0.4cm}
\end{tabular*}\medskip{}
\begin{tabular*}{12cm}{@{\extracolsep{\fill}}|>{\raggedright}p{8cm}|c|}
\hline 
\textbf{Contingency} & \textbf{17.0}\tabularnewline
\hline 
\hline 
50\% Contingency & 17.0\tabularnewline
\hline 
\end{tabular*}\medskip{}
\begin{tabular*}{12cm}{@{\extracolsep{\fill}}|>{\raggedright}p{8cm}|c|}
\hline 
\textbf{Total REDTOP} & \textbf{51.3}\tabularnewline
\hline 
\noalign{\vskip0.2cm}
\end{tabular*}\caption{\label{tab:Preliminary-cost-estimate}Preliminary cost estimate for
REDTOP under EU terms.}
\end{table}

\subsection{Cost reduction}

Along with the optimization of the detector layout, the Collaboration
is exploring alternative solutions to reduce the overall costs of
the experiment. The largest contributions to the cost of REDTOP (cfr.
Table \ref{tab:Preliminary-cost-estimate}) correspond to the sensors
for the Optical TPC and ADRIANO2, the front-end electronics and the
back-end electronics of ADRIANO2. While the number of SiPM's reading
each 10cm x 10cm tile cannot be reduced (or made smaller) without
a proportional reduction in light-yield, we are considering techniques
for ganging multiple tiles, at the cost of a reduced granularity.
This technique will reduce the number of read-out channels and the
load on the L0 trigger. Passive and active ganging is rapidly becoming
a cost-cutting resource in the latest large area detectors. The radial
granularity of ADRIANO2 could easily support ganging multiple tiles
of the same kind (i.e., glass or plastics), without loosing particle
identification power. Studies have been planned to explore the effects
that SiPM ganging would have on event pile-up, shower separation and
muon-polarization measurements. A cheaper alternative (SiPM) is being
considered for the sensors of the Optical TPC. Studies are also under
way to verify if the, intrinsically noisier, SiPM's could be a valid
replacement for the LAPPD.

\section{Conclusions}

The $\eta$ and $\eta$\textquoteright{} mesons are an excellent laboratory
for studying rare processes and physics BSM. The existing world sample
is not sufficient for breaching into decays violating conservation
laws or for searching for new particles with small coupling to the
Standard Model. A new $\eta/\eta$\textquoteright -factory is being
proposed: REDTOP. The goal is to produce \textasciitilde$10^{12}$ or
more$\eta$ mesons/yr in phase I and \textasciitilde{} 10$^{10}$
\textgreek{h}\textquoteright{} /year in phase II. Running at Fermilab
or Brookhaven would provide a factor 10$\times$ more mesons. More
running phases could use different beam species, like, for example
kaons for a \emph{stopped-K} experiment. When the future PIP-II facility
will become available, REDTOP can be upgraded into a tagged $\eta$-factory
experiment. Several laboratories have the proper infrastructures and
accelerators to host the experiment (FNAL and BNL being the most optimal).
Novel detector techniques need to be developed and/or optimized: they
would set the stage for next generation of High Intensity experiments.
The moderate cost of REDTOP (50-70 M\$), along with the many physics
measurements accessible, make this experiment very attractive.

\bibliographystyle{apsrev}
\bibliography{DPF2019_eprint}

\end{document}